\documentclass[aps,prl,twocolumn,floatfix,groupedaddress]{revtex4}

\usepackage{epsfig}
\usepackage[american]{babel}
\usepackage{amsmath}
\usepackage[dvips,usenames]{color}
\usepackage[latin2]{inputenc}
\usepackage[T1]{fontenc}
\usepackage[american]{babel}
\usepackage{color}

\usepackage{graphics}

\usepackage{amssymb, amsfonts, amsmath, amstext, txfonts}

\newcommand{\IM}{\operatorname{Im}}

\usepackage[bookmarks=true]{hyperref}

\begin{document}

\title{Itinerant and high-energy localized single-particle spectra in large-$U_d$ high-$T_c$ cuprates}

\author{O. S. Bari\v si\' c}

\affiliation{Institute of Physics, Bijeni\v cka c. 46, HR-10000, Zagreb, Croatia}

\author{S. Bari\v si\' c}

\email{sbarisic@phy.hr}

\affiliation{Department of Physics, Faculty of Science, University of Zagreb, Bijeni\v cka c. 32, HR-10000, Zagreb, Croatia}

\begin{abstract}

A theory of the underlying metallic state of large-$U_d$ high-T$_c$ cuprates is presented starting from the covalent Cu-2O 3-band model associated with a sufficiently small copper occupancy. $U_d=\infty$ is dealt by the slave fermion approach. Diagrammatic low order NCA theory in terms of the Cu-O hopping is supplemented with two independent slave particle chemical potentials which implement the $U(1)$ local gauge invariance "at average", thus avoiding the mean-field approximations. The resulting hole spectra consist of itinerant and localized states. The itinerant states close to the Fermi level are coherent. They exhibit dichotomies, the copper/oxygen in real and the nodal/antinodal in reciprocal space. The localized states, related to random Cu-O "mixed valence" fluctuations within the CuO$_2$ unit cells, are incoherent and fall well away from the Fermi level. Consequently, Luttinger's band sum rule for the conduction band is broken. Comparison with IPES, ARPES and NQR data on NCCO, LSCO, Bi2201 and Bi2212 - in order of increasing correlations - leads to remarkable agreements considering that only three bare band parameters are involved. 

\end{abstract}

\maketitle


High-T$_c$ cuprates (HTSC) represent one of the most puzzling strongly correlated electron systems ever discovered. It is widely accepted \cite{npl} that their unusual physical properties stem from large Hubbard interaction on the copper site, $U_d\sim$10 eV. The slave particle theories replace $U_d\rightarrow\infty$ by the $U(1)$ gauge invariance. Despite large $U_d$, coherent band features are experimentally well seen on the scales of 1 eV. The explanation is that large $U_d$ is "inefficient" if $n_d\approx0$ ("covalent limit" \cite{fr1}). This feature appeared early in the mean field slave particle theories (MFSPT) \cite{ktl,mr1a,mr1b} of the Emery 3-band copper-oxygen model \cite{em1}. However, the problem is that by allowing for boson condensation in the covalent regime, MFSPTs break severely \cite{mr1b} even the average $U(1)$ gauge invariance. This may not be remedied either by the approximation which includes harmonic fluctuations around the MFSPT saddle point \cite{cpl, mr1b} or by the expansions in large number $\hat n$ of Cu-spin components \cite{dch,ml1}. This has motivated us to approach \cite{bb1} the covalent regime by the low order Dyson's diagrammatic theory in terms of slave particles, which, however, develops the boson condensation. Therefore the average $U(1)$ invariance which removes the boson condensation is imposed here from the outset. At high energies such a theory is manifestly adiabatically continuous between the e- and h-doped sides \cite{lhl}.

More specifically, our perturbation approach starts from the {\it unperturbed} d$^{10}$ state ($n_d^{(0)}=0$) on copper \cite{bb1}. No spinons (bosons) at Cu sites are present in such state, while $1+x$ holes are residing on oxygens. $d^{10}$ (no hole) state is represented by one chargon (spinless fermion) on each Cu site. Such non degenerate state is thus $U(1)$ gauge invariant and belongs to the physical Hilbert subspace associated with the slave particle Hamiltonian $H$ in which the $SU(1)$ invariant charge $Q$ is equal to unity. Consequently, the number operators of bosons and $d$-fermions are equal, $n_b=n_d$. This latter relation is often called the generalized Luttinger sum rule (LSR). $H$ is also translationally invariant on the CuO$_2$ lattice. The transformation to extended dispersionless spinless fermions (characterized by $N$ quantum numbers $\vec k$) is then suitable for the development of Dyson's diagrammatic theory using $H-\lambda Q$ and the Wick theorem, both in the $\vec k$-representation. Since $H$ and the unperturbed ground state are both translationally and $U(1)$ gauge invariant, such $d=2$ theory in principle generates asymptotically the exact $U(1)$ gauge and translationally invariant ground state or breaks the symmetry in the controlled way. All the resulting physical correlations can be entirely characterized by only three parameters of the bare Emery 3-band tight-binding model, namely, by the Cu-O hybridization $t_{pd}$, the O$_x$-O$_y$ hybridization $t_{pp}$ ($<0$) and by the charge transfer gap $\Delta_{pd}=\varepsilon_p-\varepsilon_d$ ($>0$), where $\varepsilon_d$ and $\varepsilon_p$ are the Cu- and O-site energies, respectively.

In dealing with low order boson condensation we partially follow Ref.~\onlinecite{ni1} which introduces three chemical potentials, $\lambda$ for chargons (bosons), $\zeta$ for spinons (slave fermions) and $\mu$ for p-fermions, equalizing the last two. However, spinons on the Cu-sites are here taken as bosons which commute with $p$-fermions, so that such equalization is certainly not implied. Rather, the same chemical potential $\mu$ is to be shared between $p$- and $d$-fermions. In this way slave chemical potentials are taken as independent and are used to impose the average gauge invariance of diagrammatic sub-series which should become exact only asymptotically, with $\zeta^{(\infty)}=\lambda^{(\infty)}$. This is achieved through a diagrammatically founded iterative procedure for $p$-fermion, $b$-spinon and $f$-chargon single-particle propagators, with the first crucial step of which {\it chosen} to be

\begin{eqnarray}
(B^{(1)}_{\zeta^{(1)}})^{-1}&=&\omega-\varepsilon_d+\zeta^{(1)}-\beta^{(1)}_{\zeta^{(0)},\lambda^{(0)},\mu^{(1)}}\;,\nonumber\\
(F^{(1)}_{\lambda^{(1)}})^{-1}&=&\omega+\lambda^{(1)} -\phi^{(1)}_{ \lambda^{(0)}, \zeta^{(0)}, \mu^{(1)}}\;. \label{Eq001}
\end{eqnarray}

\noindent $\beta^{(1)}$ and $\phi^{(1)}$ are elementary Dyson's self-energies of spinons and chargons given by Wick's theorem \cite{bb1}. Equation~(\ref{Eq001}) includes such resummation (upgrading) \cite{bb1} that the slave self-energies in Eq.~(\ref{Eq001}) involve (beside $B^{(0)}_{\zeta^{(1)}}$ and $F^{(0)}_{\lambda^{(1)}}$) physical particle propagators hybridized through $t_{pd}$ and $t_{pp}$ (rather than through $t_{pp}$ only). For $\zeta^{(0)}=\lambda^{(0)}$ the latter coincide with the Hartree-Fock (HF) propagators \cite{bb1,lhl} of the bare 3-band Emery model characterized by the chemical potential $\mu^{(1)}$. The previous results of Refs.~\cite{bb1} are obtained by taking $\zeta^{(1)}=\lambda^{(1)}$ in Eq.~(\ref{Eq001}). The resummation acting on $t_{pd}$, $t_{pp}$-hybridized propagators in $\beta^{(1)}$ and $\phi^{(1)}$ makes the slave propagators $B^{(1)}_\zeta$ and $F^{(1)}_\lambda$ well behaving \cite{bb1} even when the $t_{pd}$-anticrossing \cite{mr1a} of the bands occurs close to the "bare" Fermi energy $\mu^{(1)}$. The boson condensation $n_b^{(1)}\rightarrow\infty$, previously obtained at $n_d^{(1)}\approx1/2$ \cite{bb1}, is removed here by allowing for the difference in slave chemical potentials $\zeta^{(1)}\neq\lambda^{(1)}$ in Eq.~(\ref{Eq001}). The latter two are determined upon imposing the average $U(1)$ invariance on spinon and chargon average numbers $n_b^{(1)}$, $n_f^{(1)}$,

\begin{eqnarray}
n_b^{(1)}=n_d^{(1)}\;,&&\; n_f^{(1)}+ n_b^{(1)}=1\;,\label{Eq002}\\
n_d^{(1)} +2n_p^{(1)}&=&1+x\;.\label{Eq003}
\end{eqnarray}

\noindent assuming $n_d^{(1)}<1$. Equations~(\ref{Eq002}), for average 
LSR $\langle n_b\rangle=\langle n_d\rangle$ and for $\langle Q\rangle=1$, are complemented by the total charge conservation for $p$- and $d$-particles in Eq.~(\ref{Eq003}). $n_d^{(1)}$ and $n_p^{(1)}$ denote the average occupation of copper and oxygen sites in the underlying bare HF state, respectively. That is, Eq.~(\ref{Eq003}) determines their common chemical potential $\mu^{(1)}$. It is important that $B^{(1)}_\zeta$ and $F^{(1)}_\lambda$ are confined \cite{ni1} and satisfy, respectively, boson and fermion commutation rules on the Cu-sites. Since $\IM B^{(1)}_\zeta$ is negative whenever finite, its retarded/advanced structure has to be specified additionally, which is provided directly by the $T=0$ perturbation theory. In this respect note that the local $F^{(1)}_\lambda$ is conjugated in time to $B^{(1)}_\zeta$. At $T=0$ this enables us to unambiguously construct the convolution 

\begin{equation}
\Sigma^{(1)}=B^{(1)}_{\zeta^{(1)}}\ast F^{(1)}_{\lambda^{(1)}}\;.\label{Eq004}
\end{equation}

Such $\Sigma^{(1)}=\Sigma^{(1)}_A+\Sigma^{(1)}_R$, which entangles spinons and chargons into the $bf$-pairs, is fermion-like and, as the local $d$-propagator should be, becomes fully fermionic when the advanced pole at $U_d+\varepsilon_d$ (state with two holes) is taken into account as well with finite spectral weight (small, for $n_d^{(1)}$ small). $\Sigma^{(1)}_A$ is the advanced pole with spectral weight $R^{(1)}$ well separated in energy from the retarded structure $\Sigma^{(1)}_R$, with the overall spectral weight $W^{(1)}$,

\begin{figure}[t]

\begin{center}{
{\includegraphics[width=.4\textwidth, angle=0]{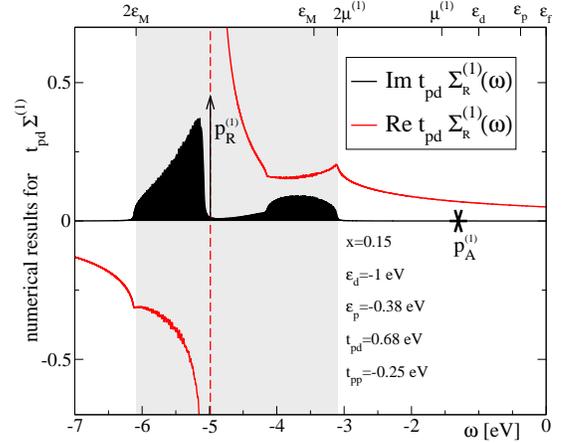}}}
\end{center} 
\caption{Numerical results for $\Sigma^{(1)}$ (in the hole picture) which sample one octant of the Brillouin zone by $N/8=1024$ $\vec k$ poles. $\Sigma^{(1)}_A$ is the advanced pole $p_A^{(1)}$ at the position $\ast$ slightly shifted from the bare value $\varepsilon_d$. Retarded structure consists of a weak split-off pole $p_R^{(1)}$ and a set of dense poles which transform into retarded cuts for $N\rightarrow\infty$. The entire retarded structure falls well away from $\mu^{(1)}$. Example is given for bare band parameters $\Delta_{pd}=0.62$ eV, $t_{pd}=0.68$ eV, $t_{pp}=-0.25$ eV, $x=0.15$.\label{Fig01}}

\end{figure}

\begin{equation}R^{(1)}=(1+\frac{1}{2}n_d^{(1)})(1-n_d^{(1)})\leq1\;,\;W^{(1)}=\frac{1}{2}(n_d^{(1)})^2<\frac{1}{2}\;.\label{Eq005}
\end{equation}

\noindent As indicated in Fig.~\ref{Fig01}, the weak "split-off" pole in $\Sigma^{(1)}_R $ is practically indistinguishable from the dense poles.

$\Sigma^{(1)}$ determines the relevant single-particle propagators $D^{(2)}$ and $P^{(2)}$. The former describes the behavior of the hole created/annihilated intermittently on Cu and the latter does the same, symmetrically on two O-sites. The pole $p_A^{(1)}$ generates together with two O-bands the {\it renormalized} coherent 3-band structure. The latter exhibit the $R^{(1)}$ decreased value of the coherent Cu-2O hybridization $t_{pd}^{(1)\;2}=R^{(1)}t_{pd}^2$, and the {\it weakly} shifted Cu-site energy $\varepsilon_d^{(1)}\approx\varepsilon_d$. Already for small $n_d^{(1)}$, the renormalized CT energy $\Delta_{pd}^{(1)}$ differs fundamentally from the MFSPT result \cite{mr1a}, although, notably, both renormalizations are then small. The renormalized HF spectral weight on Cu, $R^{(1)}z_{\vec k}^{(2D)}$, is decreased (in each of 3 bands), while $z_{\vec k}^{(2P)}=1-z_{\vec k}^{(2D)}$ on O's keeps the renormalized HF form. Further, $\Sigma_R^{(1)}$, which can be interpreted \cite{bb1} as the self-energy of the HF hybridized $p$- and $d$-particles, broadens the single-particle coherent states and generates the incoherent states occupied by holes localized in the CuO$_2$ unit cell. In cuprates they are related \cite{bb1} to the incoherent Cu-2O intracell "mixed" valence fluctuations \cite{va3, go1}. The localized states emerge {\it as soon as} $n_d^{(1)}\neq0$ and are seminal to the Mott localization within the CuO$_2$ unit cell. When/if the $R^{(1)}$-narrowed conduction band lying between $\varepsilon_d^{(1)}$ ($\Gamma$-point of the Brillouin zone) and $\varepsilon_M^{(1)}$ ($M$-point) enters the retarded continuum $\Sigma_R^{(1)}$, the immerged coherent poles are broadened by the Landau-like damping \cite{bb1}. This damping bears a close resemblance to the classical Landau damping which eventually disintegrates the plasmon when it enters the particle-hole continuum. 

Coherent spectra broadened or not, the two propagators $D^{(2)}= D^{(2coh)}+ D^{(2inc)}$ and $P^{(2)}=P^{(2coh)}+P^{(2inc)}$ determine $n_d^{(2)}$, $2n_p^{(2)}$ and through the iterated Eq.~(\ref{Eq003}) $n_d^{(2)}+2n_p^{(2)}=1+x$ fix the normalized chemical potential $\mu^{(2)}$, which thus includes the contribution of the coherent and incoherent hole-occupied states. Since the incoherent hole-occupied states fall well away from $\mu^{(2)}$, this procedure defines a {\it coherent} Fermi-line (Fs) in the $d=2$ Brillouin zone \cite{bb1}. Denoting by $1+x_{Fs}$ the normalized surface of this zone associated with hole-occupied coherent states, a deviation $x_{Fs}\neq x$ from the conventional Luttinger's band sum rule (LBSR), $x_{Fs}=x$, is obtained. This rule neglects the incoherent states, i.e., for coherent states it assumes that the doping $x_{Fs}$ uses in each band the total spectral weight equal to unity, $z_{\vec k}^{(2P)}+ z_{\vec k}^{(2D)}=1$, i.e. $R^{(1)}=1$. LBSR is broken for $R^{(1)}<1$ by two $U_d=\infty$ effects. The  incoherent hole-occupied states decrease $x_{Fs}$, while its increase stems from the reduction of the spectral weight of the coherent h-occupied states; the first one prevails for $n_d^{(1)}\ll1$ while the second takes over upon increasing $n_d^{(1)}$. It is thus predicted here that Luttinger's $x_{Fs}-x$ changes sign with $x$. 

Equations~(\ref{Eq001}-\ref{Eq005}) are meant to describe the bare parameter space $\Delta_{pd}$, $t_{pd}$, $t_{pp}$, $x$ on the covalent side of the BR crossover (phase transition in MFSP). In analogy with MFSPT \cite{mr1a} and MFfl \cite{mr1b}, we associate this crossover with the jump of $\mu^{(2)}$ into the closest O-based band, which results here in a smooth inflection (crossover) of $n_d^{(2)}$. It occurs for $n_d^{(2)}$ approaching unity from below, and, since $1>n_d^{(2)}>n_d^{(1)}$, for finite $R^{(1)}$ of Eq.~(\ref{Eq005}), i.e., for finite width of the $r=2$ conduction band. For $t_{pp}<0$, the jump of $\mu^{(2)}$ occurs for $x>0$. Although $n_d^{(2)}\approx1$ is at the verge of applicability of the present $r=2$ approximation, this suggests that the BR crossover at $x>0$ is well separated from the stabilization of the commensurate AF Mott- state close to $x\approx0$. In the present language the latter encompasses high order vertex corrections \cite{bb1} well beyond the NCA. 

On the other hand, incoherent e-occupied states appear in the $r=3$ NCA iteration of Eqs.~(\ref{Eq001}-\ref{Eq003}). That is, the $r=2,3$ incoherent states occur at energies ($\sim0.1$ eV) well away from the Fs \cite{bb1}. Remarkably, the chemical potential $\mu^{(3)}$ lies then in a ($\sim0.1$ eV) wide " high energy" window of coherent band states \cite{wtn,vnd} (de-coherence often observed by ARPES within the window at low energies ($\sim10^2$ meV) is most likely related to residual magnetic vertex corrections beyond the $r=3$ NCA). The predicted single-particle properties can be checked on the h-occupied side against the IPES data \cite{wtn, vnd, dch} while the e-occupied states are visible by ARPES. 

Comparison with experiments starts with generalizing to 3 bands the single-band expression \cite{dch,dms} for the diffuse photo-scattering intensity $I(\omega_{ph},\omega,\vec k)\propto [|M_{Cu}(\vec k,\omega_{ph})|^2 \IM D^{(r)}(\vec k,\omega)+|M_{2O}(\vec k,\omega_{ph})|^2\IM P^{(r)}(\vec k,\omega)]$, where M's are the relevant Cu and O matrix elements. 

Table~\ref{Table01} shows the bare band parameters derived by inverting the fits of the coherent $r=2$ band dispersions close to the Fs in NCCO, LSCO and Bi2201. 

\begin{table}[htb]
\begin{center}
\centering{
\begin{tabular}{r|c|c|c|c}
NCCO & $\Delta_{pd}=1$ & $t_{pd}=0.34$
& $t_{pp}=-0.34$ & $n_d^{(1)}(-0.15)=0.44$\\\hline 
LSCO & $\Delta_{pd}=0.62$ & $t_{pd}=0.68$
& $t_{pp}=-0.25$ & $n_d^{(1)}(0.15)=0.59$\\\hline 
Bi2201 & $\Delta_{pd}=1.4$ & $t_{pd}=0.35$
& $t_{pp}=-0.39$ & $n_d^{(1)}(0.15)=0.74$\\\end{tabular}
}\end{center}
\caption{Bare band parameters in eV derived for NCCO, LSCO and Bi2201. $n_d^{(1)}$ is the average HF copper occupation for a given $x$.\label{Table01}}
\end{table}

\noindent Once they are known, the $r=2,3$ high-energy de-coherences to which we turn now are fixed uniquely and differ appreciably from one material to another as discussed below. 

\begin{figure}[t]

\begin{center}{
{\includegraphics[width=.43\textwidth]{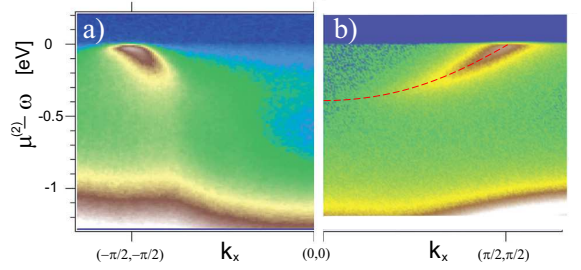}}}
\end{center}
\caption{(a) Observed ARPES intensity \cite{nco} in NCCO (using electron picture). (b) ARPES intensity predicted by the 3-band model from Table~\ref{Table01}. (a) and (b) agree upon assuming that the Cu spectral contribution dominant around the $\Gamma$-point of the conduction band at $\approx-0.5$ eV is removed from the ARPES intensity by the matrix element effect. The dashed lines is drawn as an eye guideline.\label{Fig02}} 

\end{figure}

\begin{figure}[t]

\begin{center}{
{\includegraphics[width=.35\textwidth, angle=270]{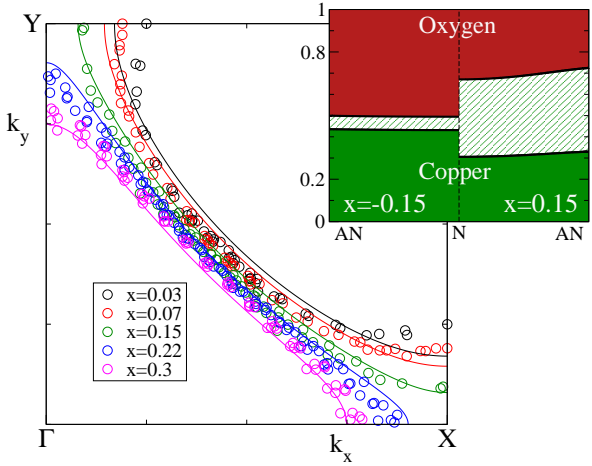}}}
\end{center}
\caption{Evolution with $x$ of the large Fermi surface in LSCO. Experimental points \cite{yo1} are obtained by averaging out the (pseudo) gap. Lines represent fits by the $r=2$ renormalized 3-band model from Table~\ref{Table01}. Inset shows the distribution of the coherent spectral weights along the Fs between antinodal (AN) and nodal (N) ranges for $x=\pm0.15$ ($x_{vH}=0.18$), O in red and Cu in green, while the depletion of the coherent Cu spectral weight corresponds to the hatched area.\label{Fig03}}
\end{figure}

 The simplest situation is encountered in the e-doped cuprates such as NCCO \cite{pan,hlm,nco,edl} shown in Fig.~\ref{Fig02}a. In this case, $n_d^{(1)}$ is small due to the overall low density limit $1+x<1$ ($x=-0.15$) in Eq.~(\ref{Eq003}). Electronic structure is then nearly coherent (see inset in Fig.~\ref{Fig03} for $x=-0.15$) and approximately described by the expression for $I(\omega_{ph},\omega^{(r)},\vec k)$ with $D^{(r)}\approx D^{(1)} $ and $P^{(r)}\approx P^{(1)}$, characterized in each band by Cu and O spectral weights $z_{\vec k}^{(1D)}$ + $z_{\vec k}^{(1P)}$=1. In particular, in the vicinity of the $\vec k=0$ $\Gamma$-point of the conduction band, the Cu spectral weight $ z_{\vec k}^{(1P)}\approx0$, so that only $M_{Cu}(\vec k,\omega_{ph})$ matters. Nevertheless, the measured $|M_{Cu}(\vec k,\omega_{ph})|^2z_{\vec k}^{(1D)}$ is small, i.e. the corresponding conduction band structure is faint \cite{pan} or turned off \cite{nco}, as illustrated in Fig.~\ref{Fig02}b around $-0.5$ eV. The only possible reason is that $M_{Cu}(\vec k,\omega_{ph})$ is then small. However, $M_{Cu}(\vec k,\omega_{ph})$ exhibits oscillations \cite{dms,mld,bnsl,jgf} versus $\omega_{ph}$ and thus possibly reappears at the $\Gamma$-point. Consistently, the oxygen component $|M_{2O}(\vec k,\omega_{ph})|^2 z_{\vec k}^{(1P)}$ of $I(\omega_{ph},\omega^{(1)},\vec k)$, which is sizeable in a rather flat \cite{mr1a} "oxygen valence band" of the 3-band model, is simultaneously well seen \cite{pan,nco} as shown in Fig.~\ref{Fig02}a. It represents a striking evidence in favor of the 3-band model \cite{em1}. 

\begin{figure}[t]

\begin{center}{
{\includegraphics[width=.45\textwidth]{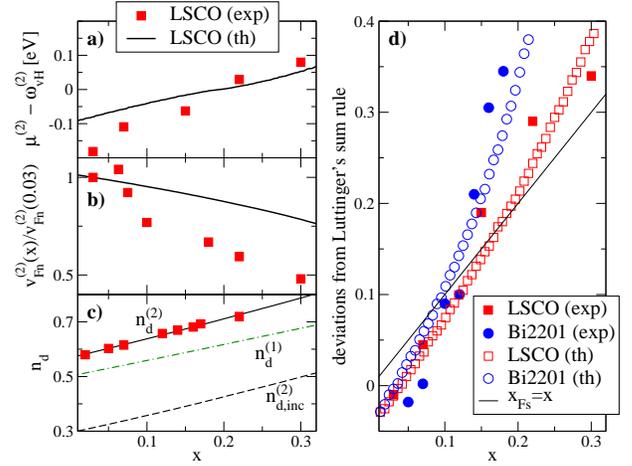}}}
\end{center}
\caption{Case of LSCO, filled squares represent experimental points while full lines are fits with three parameters of Table~\ref{Table01}. (a) Distance in eV of the Fermi level $\mu^{(2)}$ and the van Hove energy $\omega_{vH}^{(2)}$ reached for $x=x_{vH}$. (b) The same for nodal velocity $\nu_{Fn}^{(2)}(x)$ normalized by its $x=0.03$ value. (c) Total average Cu occupation $n_d^{(2)}$ and its incoherent $n_d^{(2inc)}$ component. Theoretical fit (full line) is obtained on using a liner relation (introducing \cite{ku1} two additional parameters) between the NQR resonant frequency \cite{ohs} and $n_d^{(2)}$. (d) Deviation from LBSR $x_{Fs}=x$ for LSCO and Bi2201. Filled squares are obtained from experiments, while open squares represent theoretical fits. The experimental change of sign agrees well with the prediction.\label{Fig04}}
\end{figure}

We turn next to hole doped lanthanates and Bi2201, where the renormalizations \cite{jgf,bpx,vl2,pan,mee} are presumably larger than in NCCO \cite{yo1,hs1}. Judging by the deviation \cite{yo1,hs1} $|x_{Fs}-x|$ from LBSR, lanthanates \cite{in1,yo1,hs1} at sizeable $x>0$ imply large $U_d $ renormalizations that are smaller than in Bi2201. The fit of the Fs's in lanthanates \cite{yo1} with $r=2$ (Eqs.~(\ref{Eq001}-\ref{Eq005})) using three bare parameters of Table~\ref{Table01} is remarkable as shown in Fig.~\ref{Fig03}. For $x=x^{(2)}_{vH}\approx0.18$ attainable in LSCO, $\mu^{(2)}$ crosses the X,Y antinodal van Hove points \cite{thm,mi1,hs1}, and becomes diamond- or square-like. As illustrated in Fig.~\ref{Fig03} the predicted coherent Cu-spectral weight is depleted relative to the O-component all along the Fs. The spectral weight of oxygens is (only) slightly enhanced on the nodal Fs. Such effects are also based on the Cu-O dichotomy (within the conduction band) and thus differ in essence from that found \cite{frr} in a single band Hubbard models \cite{sk9}. For $x\approx x_{vH}$, the predicted Cu and O spectral weights become comparable all along the underlying Fs of LSCO, with the Cu-component uniformly depleted by $R^{(1)}$ of Eq.~(\ref{Eq005}). This reduces in particular the $t_{pd}$ induced $X,Y$ vH singularity at $\omega=\omega^{(2)}_{vH}$ but, nevertheless, the chemical potential $\mu^{(2)}$ is "pinned" to this singularity \cite{hrm}, $\partial\mu^{(2)}/\partial x=0$ for $\omega=\omega_{vH}^{(2)}$. The predicted $x$-behavior of $\mu^{(2)}-\omega_{vH}^{(2)}$ is compared to measurements in Fig.~\ref{Fig04}b (note that uniform Madelung shifts cancel out from $\mu^{(2)}-\omega_{vH}^{(2)}$ \cite{ku1}). Moreover, the predicted $x_{Fs}-x$ vs. $x$ in Fig.~\ref{Fig04}d is found to change sign in agreement with measurements \cite{yo1,hs1}. In Fig.~\ref{Fig04}c we also plot the underlying nodal velocity $\nu_{Fn}^{(2)}$ taken just beyond the "low energy kink" \cite{zhu}. Its decrease at large $x$ is faster than in the corresponding bare 3-band model and the rate of change of $\nu_{Fn}^{(2)}$ agrees reasonably well with observations. Such drop of $\nu_{Fn}^{(2)}$, accompanied in Fig.~\ref{Fig04}d by the steady increase of $n_d^{(2)}$ vs. $x$, agrees with the NQR/NMR measurements \cite{ohs,haa} as well. $\partial n_d/\partial x\approx1/3$ obtained \cite{ku1} for $x\geq0$ shows that covalent lanthanates (as well as other similar cuprates \cite{ohs,tg1,ku1,haa}) fall well away from the BR crossover. The results agree qualitatively with a quite recent ARPES observation \cite{jcg2} of the band broadening on optimally to overdoped LSCO, which exhibits a sharp coherence threshold at 0.1-0.2 eV. 

Bi2201 provides a similar example corresponding to a bit stronger renormalizations \cite{hs1}. Indeed, after changing sign, the deviation $x_{Fs}>x$ from LBSR becomes relatively large \cite{hs1} already for $0.15<x<0.2$. However, as predicted by Eqs.~(\ref{Eq001}-\ref{Eq005}), the renormalized CT scale $\Delta_{pd}^{(1)}$ is of the same order of magnitude in Bi2201 as in NCCO and LSCO, although $R^{(1)}$ is smaller according to Table~\ref{Table01}. The Bi2201 photo-generation from the $\Gamma$-point, which in LSCO and Bi2201 (like in NCCO) extrapolates to the binding energy $\sim-0.5$ eV, is invisible \cite{jgf,bpx,vl2,jcg,mee,mts} or faint \cite{pan}. The likely explanation is that it is suppressed by the Cu-matrix element $M_{Cu}(\omega_{ph})\approx0$, in analogy with NCCO. Notably, in this respect the $r=3$ coherent picture holds around the $\Gamma$-point even when the latter is imbedded in the continuum of localized chargon-spinon pairs, since in this case the effective coupling between itinerant and localized states vanishes \cite{bb1}. 

In Bi2212, the photo-generation from the conduction band $\Gamma$-point at $\sim-0.5$ eV, appears together with incoherent "waterfalls" \cite{kdy, ins, jgf} at high energies when photon energy $\omega_{ph}$ is varied. This was originally attributed \cite{kdy,ins} to the oscillatory behavior of the matrix element $M_{Cu}(\omega_{ph})\neq0$, while here, the predicted ARPES intensity is a joint effect of $M_{Cu}(\omega_{ph})$ and coherent or incoherent many-body features of  $D^{(3)}$ and/or $P^{(3)}$ propagators. The comparison with NCCO \cite{nco,edl}, by keeping the spirit of Figs.~\ref{Fig02}, suggests that Bi2212 is covalent but falls closer to the BR crossover, which, however, is not yet reached, even with overdoping. Due to sizeable $n_d$, the analysis of the latter regime requires further quantitative study, which will be presented elsewhere.

Let us finally mention that the particle-hole convolutions of NCA $P^{(2)}$ and $D^{(2)}$ are seminal to the Fermi (FL) and spin (SL) liquids proposed in Refs.~\onlinecite{pns} considering the $T$-dependent multi-component uniform magnetic susceptibility, and supported by recent observations \cite{haa} for $x>0$. SL consists of spin/charge separated states in which only spin-flips are possibly traveling in presence of kinematical two-particle (four leg) interactions such as $p-p$ repulsion \cite{bb1,sk7}, $b-b$ super-exchange and RKKY, and mixed $f-b$ interaction \cite{bb1}. In particular, the mixed $f-p$ vertex couples the spin-flips on Cu- and O-sites. In contrast, only FL is directly involved in the dc conductivity due to the formation of nodal arcs. The high order vertex corrections are then less important for the low energy resistivity than for spin susceptibility and the $r=2,3$ NCA in strongly correlated cuprates is consistent with the use of the nodal FL with low order kinematical Umklapp corrections \cite{nvn,go7,ku3,ric}. These important issues will be discussed in more detail elsewhere.

\begin{acknowledgments}

We gratefully acknowledge enlightening discussions and correspondences with J. Friedel, L. P. Gor'kov, M. Greven, A. Fujimori, A. J. Millis, and D. Pines, with our colleagues and collaborators I. Kup\v ci\' c, D. K. Sunko and M. \v{S}unji\'{c}, as well as with N. Bari\v{s}i\'{c}. This work was supported by the Croatian Government under Projects $119-1191458-0512$ and $035-0000000-3187$.

\end{acknowledgments}

\end{document}